\documentclass[sigconf]{acmart}
\usepackage{multirow}
\usepackage{multicol}
\AtBeginDocument{%
  \providecommand\BibTeX{{%
    \normalfont B\kern-0.5em{\scshape i\kern-0.25em b}\kern-0.8em\TeX}}}

\copyrightyear{2021} 
\acmYear{2021} 
\setcopyright{rightsretained} 
\acmConference[ICONS 2021]{International Conference on Neuromorphic Systems 2021}{July 27--29, 2021}{Knoxville, TN, USA}
\acmBooktitle{International Conference on Neuromorphic Systems 2021 (ICONS 2021), July 27--29, 2021, Knoxville, TN, USA}
\acmDOI{10.1145/3477145.3477267}
\acmISBN{978-1-4503-8691-3/21/07}

\acmConference[ICONS 2021]{International Conference on Neuromorphic Systems 2021}{July 27--29, 2021}{Knoxville, TN, USA}
\acmBooktitle{International Conference on Neuromorphic Systems 2021 (ICONS 2021), July 27--29, 2021, Knoxville, TN, USA}
\acmISBN{978-1-4503-8691-3/21/07}

\begin{document}

\title{{Signals to Spikes for Neuromorphic Regulated Reservoir Computing and EMG Hand Gesture Recognition
}}
\author{Nikhil Garg}

\affiliation{%
  \institution{3IT - LN2 - CNRS UMI-3463
Université de Sherbrooke
}
  \country{Canada}}
  
\affiliation{%
  \institution{BITS Pilani - Goa Campus
}
  \country{India}}  
  
\email{nikhil.garg@usherbrooke.ca}

\author{Ismael Balafrej}
\affiliation{%
  \institution{NECOTIS – 3IT \\
Université de Sherbrooke
}
  \country{Canada}}
\email{ismael.balafrej@usherbrooke.ca}

\author{Yann Beilliard}
\affiliation{%
  \institution{3IT - LN2 – CNRS UMI-3463
Université de Sherbrooke
}
  \country{Canada}}
\email{yann.beilliard@usherbrooke.ca}

\author{Dominique Drouin}
\affiliation{%
  \institution{3IT - LN2 – CNRS UMI-3463
Université de Sherbrooke
}
  \country{Canada}}
\email{dominique.drouin@usherbrooke.ca }

\author{Fabien Alibart}
\affiliation{%
  \institution{3IT - LN2 – CNRS UMI-3463
Université de Sherbrooke
}
  \country{Canada}}
  
  \affiliation{%
  \institution{IEMN, Université de Lille
}
  \country{France}}
  
 \email{fabien.alibart@usherbrooke.ca}

\author{Jean Rouat}
\affiliation{%
  \institution{NECOTIS - 3IT \\
Université de Sherbrooke
}
  \country{Canada}}
\email{jean.rouat@usherbrooke.ca}

\renewcommand{\shortauthors}{Garg et al.}

\begin{abstract}
Surface electromyogram (sEMG) signals result from muscle movement and hence they are an ideal candidate for benchmarking event-driven sensing and computing.
We propose a simple yet novel approach for optimizing the spike encoding algorithm's hyper-parameters inspired by the readout layer concept in reservoir computing. Using a simple machine learning algorithm after spike encoding, we report performance higher than the state-of-the-art spiking neural networks on two open-source datasets for hand gesture recognition. 
The spike encoded data is processed through a spiking reservoir with a biologically inspired topology and neuron model. 
When trained with the unsupervised activity regulation 
CRITICAL algorithm to operate at the edge of chaos, the reservoir yields better performance than state-of-the-art convolutional neural networks. The reservoir performance with regulated activity was found to be 89.72\% for the Roshambo EMG dataset and 70.6\% for the EMG subset of sensor fusion dataset. Therefore, the biologically-inspired computing paradigm, which is known for being power efficient, also proves to have a great potential when compared with conventional AI algorithms. 
  
\end{abstract}




\maketitle

\section{Introduction}

Worldwide, there are millions of people with severe upper limb amputations \cite{Kannenberg2017ActiveUP} that require prosthesis implants to perform routine activities \cite{lusardi2013orthotics, cordella2016literature}.  Surface electromyography (sEMG) is a technique that measures the electrical activity in response to the muscle's stimulations by a nerve and is a popular method to sense the users intention and drive the prosthesis due to its non-invasive nature and low cost \cite{Scheme2011ElectromyogramPR, Phinyomark2011ARO}. Constrained by low-power budget, processing these signals in real time is a challenging task. Neuromorphic computing \cite{Mead2020} is about rethinking algorithms, circuits, and devices to match the principles of computation governing the human brain.
It might be a viable solution to EMG analysis and classification.
Asynchronous circuits, local learning rules, analog computation, spike-based information representation, and processing are key features of the neuromorphic computing paradigm that distinguish it from classical digital computers. The networks for neuromorphic computing is referred to as spiking neural networks.\\

Third-generation neural networks, commonly referred to as Spiking Neural Networks (SNNs), perform computation in a more biologically plausible way. In a SNN, the information is represented and transmitted in discrete spikes of fixed amplitude, and thus such a network requires input to be of the same form. Dynamic vision sensor or silicon retina \cite{Lichtsteiner2008DVSarray} for capturing visual information, and artificial cochlea \cite{Wen2009ASC} for sensing audio signals are some of the well-established event-based sensors. While event-based sensors for biomedical signal recording instruments were conceptually proposed in the past \cite{Corradi2015BMI}, they are still not available commercially. Therefore the recorded EMG data must be converted to spike trains for feeding the SNN. It is also worth mentioning that this conversion should occur in a lossless manner for benchmarking the performance of SNNs. In the temporal contrast method \cite{Petro2020encoding}, the generation of events is led by a change in signal and not by steady signals as done in rate-based encoding methods \cite{Chen2006AsynchronousBP}, and energy is not consumed in the absence of signal activity. The spike converted data is processed through an efficient SNN architecture.\\

Optimal network topology and plasticity mechanisms are critical for the functionality of a SNN. Reservoir computer \cite{Maass2002RealTimeCW} is used for maximizing spikes dynamics. Reservoir computing uses both spatial and temporal features from signals and projects them on a higher dimensional vector space and therefore seems to be a powerful method to compute spatiotemporal patterns of activities. The performance of a reservoir is highly dependent on the topology, and synaptic weights. Therefore an efficient synaptic plasticity algorithm is required for the reservoir to operate efficiently. It is believed that dynamic systems function more efficiently as computers at the "edge of chaos".
The 
Lyapunov exponent quantifies the rate of separation of close trajectories. The network activity dies down quickly in the case of a negative 
Lyapunov exponent and explodes in a positive 
Lyapunov exponent. The edge of chaos occurs when "the largest 
Lyapunov exponent transitions from negative to positive" \cite{Legenstein2007EdgeOC, Carroll2019MutualIA}. In this study, we use the CRITICAL \cite{CRITICAL2012} plasticity rule to adjust the reservoir's recurrent connection weights and drive the network towards the edge of chaos.\\ 

In the past, reservoirs with fixed weights have been used to recognize hand gestures using forearm EMG \cite{Ma2020AICAS, Ma2020JETCAS}.
Another popular method of processing EMG signals is features extractions \cite{Spiewak_2018} and machine learning, but this requires tailoring the features \cite{Peng2015} based on the problem at hand and the training is not end-to-end. Deep learning \cite{Phinyomark_2018} methodologies, and non-spiking reservoir computing \cite{Bostrm2013ModelFA} have also been studied in the past for EMG signal processing, but such architectures are power-hungry and difficult to deploy on edge in real-time; moreover, the performance of deep neural networks is limited by the availability of large datasets. While there are studies done in the past to evaluate the performance of plastic spiking reservoirs for processing electroencephalogram (EEG) signals \cite{Moinnereau2018ClassificationOA}, such a study of EMG signal processing is not known to the authors.
In this study, we wish to evaluate the impact of synaptic plasticity in recurrent connections on a reservoir performance.\\ 

In this work, we first discuss the sEMG dataset and spike encoding algorithm. To this end, we propose a novel yet straightforward algorithm to evaluate the encoding efficiency for the classification task. Subsequently, we simulate several different reservoir computers and ask whether the best performance is achieved on this edge of chaos. We use the CRITICAL \cite{CRITICAL2012} branching factor auto-regulation algorithm for driving the network to this state. We then evaluate multiple ingredients of a reservoir of spiking neurons (feeding input to the reservoir, properties of the neurons, connection density, and plasticity) and propose strategies for designing an efficient reservoir computing system. We report performance better than the present state of the art for sEMG-based gesture recognition for \cite{donati_elisa_2019_3194792, ceolini_enea_2020_3663616} in the results section.

\section{Methods}

\subsection{Dataset}

We use two publicly available hand gesture recognition datasets collected through the Myo armband. The armband has eight equally spaced sensors with a sampling rate of 200 Hz. 

The Roshambo dataset \cite{donati_elisa_2019_3194792} consists of 3 gestures, collected from 10 participants, across 3 sessions. Each session comprises 5 trials for each gesture; hence there is 450 trials in the entire dataset. The initial and final 600 ms of each trial were trimmed off as done in \cite{Ma2020AICAS} and \cite{Ma2020JETCAS} as to benchmark the performance equitably. The EMG subset of sensor fusion dataset, also referred to as the 5 class dataset \cite{ceolini_enea_2020_3663616}, consists of 5 gestures. The dataset comprises recordings from 21 participants. Each participant took part in 3 sessions, and each session comprised five trials of each gesture. Each trial is for 2 seconds. Hence, there was 1575 trials in the entire dataset. As there are three sessions, two of them are selected for training and one is used for testing with the aforementioned approaches. Hence, a three-fold cross-validation accuracy was computed for all the accuracies in this work.

\begin{figure*}[h]
  \centering
  \includegraphics[width=\linewidth]{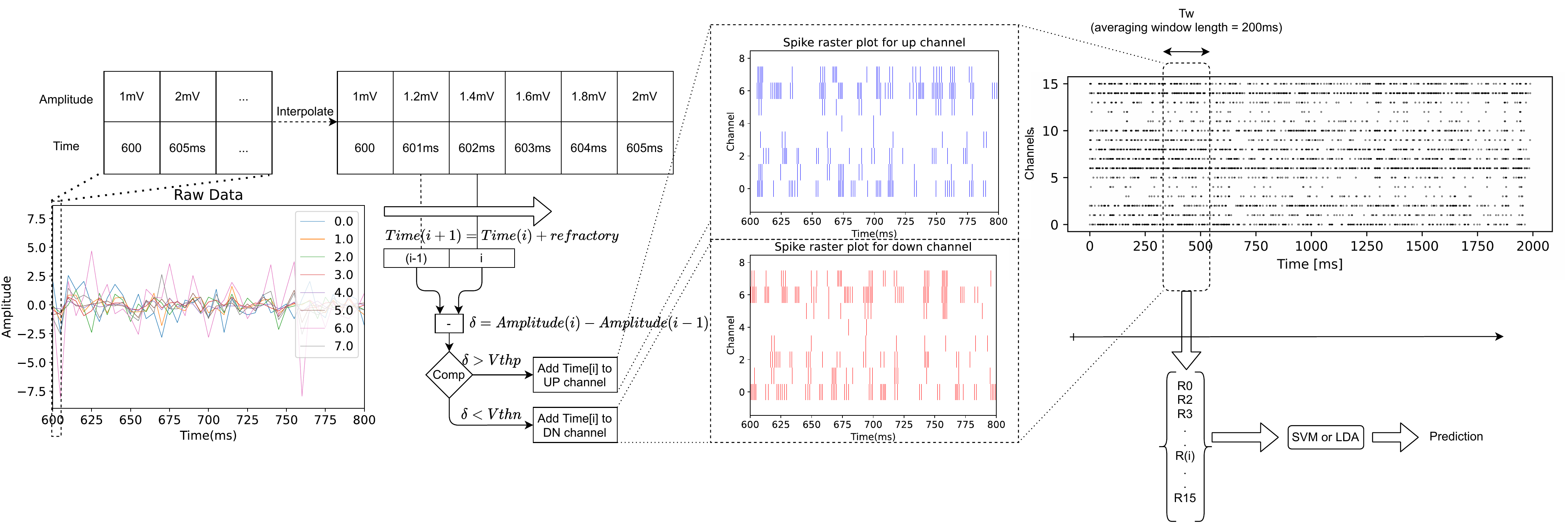}    

  \caption{Schematic representation of the 
proposed spike encoding and evaluation 
baseline. 
The eight continuous analog EMG signals are encoded as UP and down (DN) spikes. The data is first interpolated to increase the temporal resolution 
like in~\cite{Donati2019a}.
The difference between successive amplitudes is compared to a positive and a negative threshold to obtain spike times. Eight UP (blue) and eight DN (red) resulted in 16 channels. The spike counts is then computed across a time window of 200 ms duration for each of the 16 spike trains and fed into an SVM classifier. }
  \Description{The eight continuous analog EMG signals are encoded as UP and down (DN) spikes. The data is first interpolated and the difference between successive amplitudes is compared to a positive and a negative threshold to obtain spike times. The average firing rate is computed across a time window of 200 ms for each of the 16 spike trains and fed into an SVM classifier.}
  \label{fig:pipeline_without_reservoir}
\end{figure*}

\subsubsection{Encoding}

To realize the full potential of event-driven computing, it is essential to perform event-driven sensing. We use the temporal contrast method \cite{Lichtsteiner2008DVSarray, Ma2020AICAS, Ceolini2020Frontiers} to encode the continuous analog signal to discrete spike trains. In this method, the variation in amplitude is calculated across time. If the difference between a successive sample exceeds a predefined threshold, a spike is generated. We use two separate channels to better represent the signal in a lossless fashion for excitation (positive change in signal amplitude) or depression (negative change in signal amplitude) as shown in Figure \ref{fig:pipeline_without_reservoir}. For hardware implementation, the successive difference would be calculated with the help of an optimal amplifier topology \cite{Jimnez2017AMF}, and subsequently, this difference would be compared with a predefined threshold with the help of a comparator. This approach's key parameters are the positive threshold (Vthp), negative threshold (Vthn), interpolation factor, and refractory period. A single analog channel results in two segments of spike times (UP and DN). 




\subsubsection{Evaluating the encoding algorithm}

For solving the classification problem, every algorithm applied to the 
signal is aimed at extracting information that is valuable for class differentiation. In the literature, the conventional 
metric used for evaluating the encoding efficiency is the ability to reconstruct the original analog signal and minimize the mean square error (MSE) between the original and reconstructed signals \cite{Petro2020encoding}. The above is based on the assumption that the entire spectrum of the signal is of interest and any information loss has to be minimized. However, much of the signal component might not be of interest for classification. Tailoring the encoding hyper-parameters to reconstruct the background signal defeats the purpose of event-based sensing. While some studies \cite{Guo2021NeuralCI, Anumula_2018} evaluate the SNN performance to optimize the encoding hyperparameters, the performance is effected by the SNN topology and learning algorithm. \\

To this end, we propose a novel approach to evaluate and tune our signals to spikes encoding front ends in the context of classification tasks. We refer to as \emph{spike encoding \& evaluation baseline} (\emph{spike enc. eval. baseline)} (as illustrated in Figure~\ref{fig:pipeline_without_reservoir}).
This is, in turn, inspired by the readout method, which is a popular classification strategy for inferring the states of reservoirs. 
As shown in Figure \ref{fig:pipeline_without_reservoir}, rate vectors averaged over 200 ms window are classified using a machine learning classifier (linear discriminant analysis (LDA) or support vector machine (SVM)). With help of grid search, the accuracy was computed for different values of positive (0,1) and negative (-1,0) threshold. The selected ones are positive and negative threshold as 0.5 V and -0.5 V respectively. 
\subsection{Spiking neural network}

\subsubsection{Recurrent neural networks and Reservoir computing}

Recurrent neural networks incorporate the past input memory for computing the output. Recurrent neural networks are thus helpful for processing time-domain signals. 
Reservoirs 
inherently behave as recurrent neural networks and are also referred to as Liquid State Machines (LSM)~\cite{Maass2002RealTimeCW} and Echo State Networks (ESN)~\cite{Jaeger2001TheechoST}.
An efficient reservoir projects the input signal into a space of linearly separable vectors so that the decision can be computed by processing the network state (e.g. firing rate) through simple machine learning classifiers. Inferring from the reservoir of neurons consists of feeding in a signal of interest and matching the reservoir neurons' time-series signals to a training signal related to the input signal of the same class \cite{Carroll2019MutualIA}.

\subsubsection{Neuron}

Neurons perform temporal and spatial integration of weighted spikes from the synapses and add non-linearity to the system. Spiking neural networks are composed of neuron models inspired by the biological neuron. Leaky integrate and fire (LIF) neuron \cite{Abbott1999LapicquesIO} is a good balance between biological plausibility and computational feasibility and hence used in this study. A key feature of a biological neuron is homeostasis or adaptation; when a neuron is stimulated repeatedly, it becomes less sensitive to the stimulation, and the firing rate decreases. To account for homeostasis, we implement adaptive LIF neurons \cite{CRITICAL2012}.  

The neuron integrates the incoming signal, with leakage (time constant of $\tau$ to resting membrane voltage of $v_{0}$), which is described by 

\begin{equation}\label{neuron_eq_1}
    \frac{\mathrm{d} v_{}}{\mathrm{d} t} = \frac{-(v-v_{0})}{\tau}
\end{equation}

When the neuron membrane potential ($v$) reaches a threshold level ($v_{th}$), a spike is generated and the neuron membrane potential resets to the reset voltage ($v_{0}$). After each spike, the threshold is incremented by an adaptive threshold increment value ($v_{thi}$), which in turn also leaks (to $v_{th0}$ with time constant of $\tau_{v_{th}}$ ) to decrease the threshold in the absence of stimulation which is described by 

\begin{equation}\label{neuron_eq_2}
    v_{th} = v_{th}+v_{thi} 
\end{equation}

\begin{equation}\label{neuron_eq_3}
    \frac{\mathrm{d} v_{th}}{\mathrm{d} t} = \frac{-(v_{th}-v_{th0})}{\tau_{v_{th}}}
\end{equation}

The parameters along with descriptions are tabulated in Table \ref{tab:neuron_params}. 

\begin{figure*}[h]
  \centering
  \includegraphics[width=\linewidth]{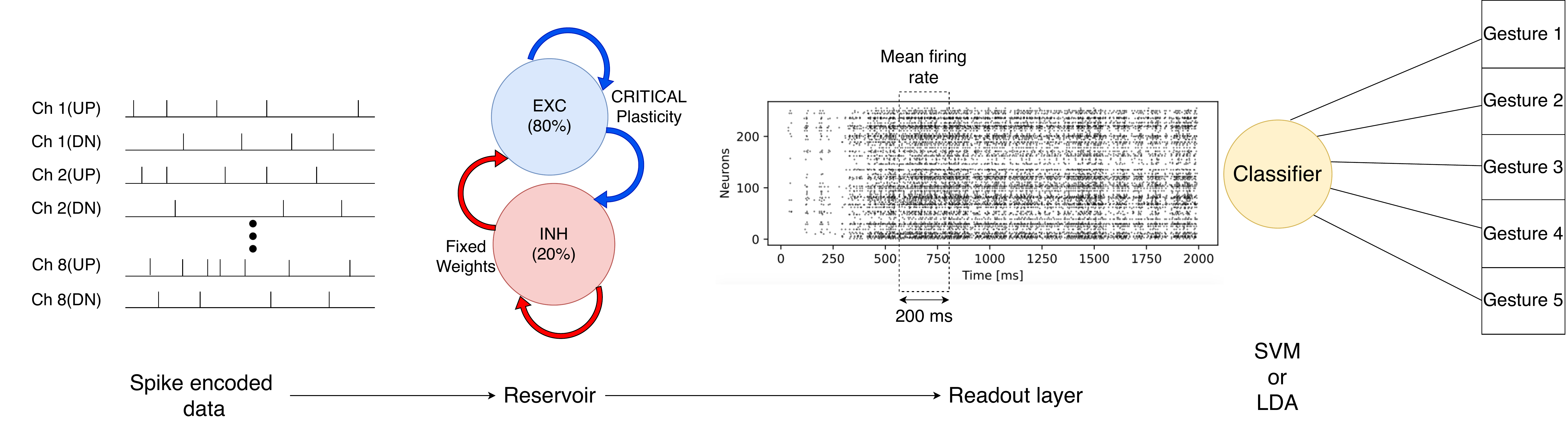}
  
  \caption{Pipeline for EMG processing with a plastic reservoir. The spike encoded data from 8 electrodes consisting of 16 channels is fed to the reservoir of spiking neurons with 80\% excitatory neurons and 20\% inhibitory neurons. The input connections are made invariable of UP/DN channel or exc/inh neuron type. CRITICAL plasticity is implemented on the excitatory connections (in blue), while the weights of inhibitory connections (in red) are fixed. The average spike rate vector of reservoir neurons over a 200 ms window is classified using an SVM or LDA-based readout classifier to 5 gestures for sensor fusion dataset and 3 gestures for Roshambo dataset }
  \Description{EMG processing pipeline and reservoir topology}
  \label{fig:pipeline_with_reservoir}
\end{figure*}

\begin{table}[]
\centering
\begin{tabular}{lll}
\hline
Parameter  & Value & Description                                                                                 \\ \hline
v$_{0}$         & 0 V         & \begin{tabular}[c]{@{}l@{}}Resting potential\\ Reset potential\end{tabular}           \\ \hline
$\tau$        & 15 to 25 ms  & Leak time constant                                                                    \\ \hline
refractory & 1 ms         & Refractory period                                                                     \\ \hline
$v_{th0}$        & 1 V         & \begin{tabular}[c]{@{}l@{}}Adaptive threshold \\ Initial and leak value\end{tabular} \\ \hline
$v_{thi}$        & 100 mV      & \begin{tabular}[c]{@{}l@{}}Adaptive threshold \\ Increment\end{tabular}               \\ \hline
$\tau_{v_{th}}$     & 50 ms       & \begin{tabular}[c]{@{}l@{}}Adaptive threshold\\ Leak time constant\end{tabular}       \\ \hline
\end{tabular}
\caption{Neuron parameters}
\label{tab:neuron_params}
\end{table}

\subsubsection{Synapses and activity regulation}

The synaptic transmission is based on equation (\ref{syn_eq_1}). The weights are initialized in the range [0, 0.25] with uniform distribution. Plasticity is implemented in \textit{'w'} governed by the CRITICAL learning rule \cite{CRITICAL2012} which maximizes the reservoir's sensitivity. 
\begin{equation}\label{syn_eq_1}
    v_{post} = v_{post}+w 
\end{equation}

\begin{table}[]
\begin{tabular}{ll}
\hline
Parameter & Value \\ \hline
Shape of macrocolumn & {[}2,5,1{]} \\ \hline
Shape of minicolumn & {[}4,4,2{]} \\ \hline
Connectivity & Small world \\ \hline
Size of reservoir population & 320 \\ \hline
\begin{tabular}[c]{@{}l@{}}Proportion of excitatory neuron\\  population\end{tabular} & 0.8 \\ \hline

\begin{tabular}[c]{@{}l@{}}Number of \\  recurrent connections\end{tabular} & 1161 \\ \hline
Number of input connections & 174 \\ \hline
Initial input weights & 0 to 1 \\ \hline
Initial reservoir weights & 0 to 0.25 \\ \hline
\begin{tabular}[c]{@{}l@{}}Learning rate of  CRITICAL\\  learning rule\end{tabular} & 0.1 \\ \hline
\begin{tabular}[c]{@{}l@{}}Target branching factor for CRITICAL\\ learning rule\end{tabular} & 1 \\ \hline
\begin{tabular}[c]{@{}l@{}}Algorithm to adapt weights of recurrent\\ connections\end{tabular} & CRITICAL \\ \hline
\end{tabular}
\caption{Parameters for a reservoir of 320 neurons}
\label{tab:reservoir_params}
\end{table}
\raggedbottom

Reservoirs are prone to various problems in the balance of the spike activity. Either too much or too little activity is not effective in maintaining a working memory inside the reservoir, which is crucial for their implementation. The "edge-of-chaos," or criticality, has been shown to be an adequate target for reaching this balance. In order to maintain this balance, an adaptation mechanism must be implemented inside the reservoir. For this purpose, the CRITICAL learning rule \cite{CRITICAL2012} was chosen as a local weight regulation plasticity rule. The usage of CRITICAL should allow the weights of the reservoir to be in a more desirable state than fully random initialization. The CRITICAL learning rule tunes the weights such that the local branching factor of every neuron reaches a target value. It is believed that maintaining the local branching factor near one will keep the reservoir at this "edge-of-chaos" \cite{Haldeman2005-pa}. A branching factor of 1 conceptually means that a spike generated by any reservoir neuron would lead to, on average, a single spike in all of the postsynaptic neurons.

\vspace{-2.3mm}

\subsubsection{Topology}

\begin{figure}[h]
  \centering
  \includegraphics[width=\linewidth]{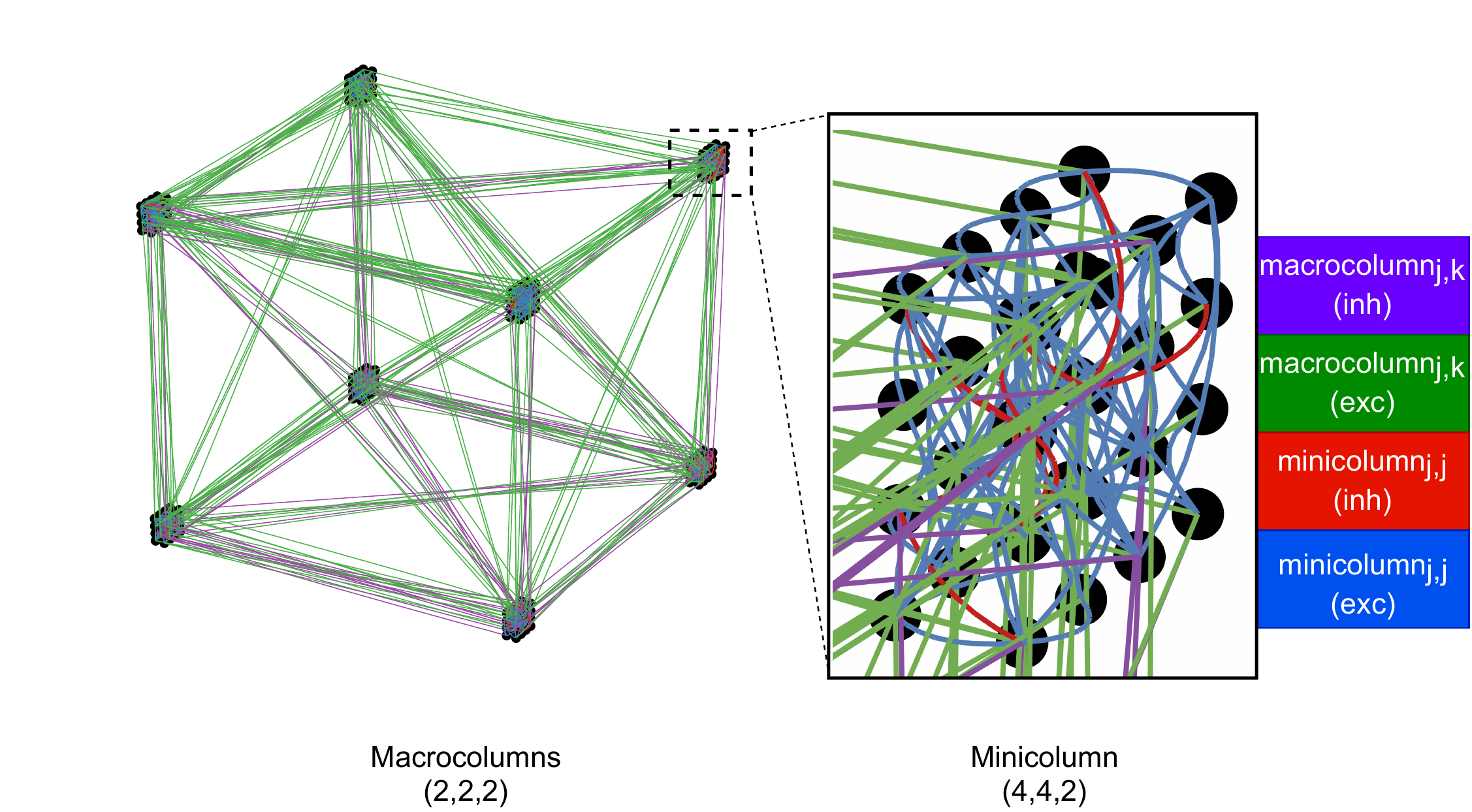}
  \caption{
Example of a small world topology for a reservoir  of macrocolumn shape : [2,2,2], minicolumn shape of [4,4,2],  and 256 neurons. The excitatory and inhibitory connections between neurons from the same minicolumn and across macro columns are shown in the legend's lines of color. }
  \Description{Small world topology for the reservoir of macrocolumn shape : [2,2,2], minicolumn shape of [4,4,2], and 256 neurons. The excitatory and inhibitory connections between neurons from the same minicolumn and across macro columns are shown in the legend's lines of color.}
  \label{fig:reservoir}
\end{figure}

It is important to note that CRITICAL does not influence the reservoir's connectivity, but only the weights. Therefore, the network connectivity must still be chosen adequately. For this purpose, the initial connectivity of the reservoir is a small-world-like topology \cite{Wijesinghe2019, Davey2006HighCS} with the same parameters that were optimized with biologically realistic spectral values in \cite{Balafrej2020PCRITICALAR}. The neurons are arranged in tri-dimensional minicolumns, as shown in Figure \ref{fig:reservoir}, and the probability of connection between two random neurons is purely euclidean distance-based. As an example, in Figure~\ref{fig:reservoir}, there is 8 minicolumns in the macrocolumnar organization, with each having 32 neurons for a total of 256 neurons inside the reservoir. The connections between the input neurons and the reservoir neurons are made randomly, with the total number of input connections being 15\% of reservoir connections \cite{CRITICAL2012}. The input connection density was optimized for performance. Lastly, all of the reservoir neurons are read when feeding the classifier.

All parameters for a reservoir of 320 neurons are listed in Table \ref{tab:reservoir_params}. The number of neuron were chosen to 320 in order to benchmark the results with respect to the ones reported in \cite{Ma2020AICAS, Ma2020JETCAS}. The simulations were performed with time step, \textit{dt = 1 ms} and using Brian2 \cite{Stimberg2019} simulator. 

\subsection{Readout}

The state of the reservoir is extracted in the form of each neuron's mean firing rate across a time duration of 200 ms. 
The readout layer is inherently memoryless, as its value is not dependent upon a signal from the past. The resultant rate vector is fed into a classical machine learning classifier, which is a support vector machine with a radial basis function kernel \cite{Boser_1992} in this study. The regularization parameter (C) was chosen to be 1 while the parameter kernel coefficient (Gamma) was chosen to be 1/(number of features), note the number of features is equal to 16 for spike encoding baseline and number of reservoir neurons for reservoir computing. 

\begin{table}[]
\begin{tabular}{llll}
\hline
Dataset & Method & \begin{tabular}[c]{@{}l@{}}Readout \\ classifier\end{tabular} & Accuracy ($\mu \pm \sigma$) \\ \hline
\multirow{6}{*}{Roshambo} & \multirow{2}{*}{\begin{tabular}[c]{@{}l@{}}Spike Enc. \&\\ Eval. Baseline \end{tabular}} & LDA & 74.61 $\pm$ 2.44\% \\ \cline{3-4} 
 &  & SVM & 85.44 $\pm$ 0.75\% \\ \cline{2-4} 
 & \multirow{2}{*}{\begin{tabular}[c]{@{}l@{}}Reservoir\\ (No plasticity)\end{tabular}} & LDA & 76.44 $\pm$ 1.80\% \\ \cline{3-4} 
 &  & SVM & 81.83 $\pm$ 0.93\% \\ \cline{2-4} 
 & \multirow{2}{*}{\begin{tabular}[c]{@{}l@{}}Reservoir\\ (CRITICAL)\end{tabular}} & LDA & 83.06 $\pm$ 0.92\% \\ \cline{3-4} 
 &  & SVM &\textbf{88.00 $\pm$ 0.29\%} \\ \hline
\multirow{6}{*}{\begin{tabular}[c]{@{}l@{}}Sensor \\ fusion\\ (EMG)\end{tabular}} & \multirow{2}{*}{\begin{tabular}[c]{@{}l@{}}Spike Enc. \&\\ Eval. Baseline \end{tabular}} & LDA & 53.03 $\pm$ 1.38\% \\ \cline{3-4} 
 &  & SVM & 65.18 $\pm$ 3.28\% \\ \cline{2-4} 
 & \multirow{2}{*}{\begin{tabular}[c]{@{}l@{}}Reservoir\\ (No plasticity)\end{tabular}} & LDA & 55.97 $\pm$ 2.07\% \\ \cline{3-4} 
 &  & SVM & 62.33 $\pm$ 2.05\% \\ \cline{2-4} 
 & \multirow{2}{*}{\begin{tabular}[c]{@{}l@{}}Reservoir\\ (CRITICAL)\end{tabular}} & LDA & 61.96 $\pm$ 2.99\% \\ \cline{3-4} 
 &  & SVM & \textbf{70.60 $\pm$ 3.65\%} \\ \hline
\end{tabular}
\caption{Comparison of 3-fold cross-validation accuracies obtained with different readout methods (LDA and SVM with a radial basis function (RBF) kernel) for Roshambo dataset (3 class), and sensor fusion dataset (5 class) performed on encoded data, reservoir without plasticity and reservoir with CRITICAL plasticity. Note that we use reservoir with 320 neurons for 3 class problem, and 2048 neurons for 5 class problem, with a topology initialized from a small-world distribution with parameters coming from the optimization of a biorealistic eigenspectrum, as done in \cite{Balafrej2020PCRITICALAR}.}
\label{tab:classifier_comp}
\end{table}

\section{Results and Discussions}

We propose two pipelines, 
the Spike Encoding \& Evaluation Baseline as presented in Figure~\ref{fig:pipeline_without_reservoir}, and an encoder followed by a CRITICAL-enabled reservoir as shown in Figure \ref{fig:pipeline_with_reservoir}. Rather than minimizing the error in signal reconstruction, we optimize the encoding parameters by maximizing the accuracy of a supervised readout layer trained with the average rate vectors as shown in Figure  \ref{fig:pipeline_without_reservoir}.  Data from 1 session was taken for testing from the three sessions, and the other two were used for training the machine learning classifier. The three-fold classification accuracy for two machine learning classifiers : LDA and SVM (RBF kernel) are reported in Table \ref{tab:classifier_comp}. We observe a higher accuracy for the SVM classifier than the LDA classifier and hence use that for comparing performance with literature (Table \ref{tab:benchmark_lit}). 

With the Spike Encoder \& Evaluation Baseline we obtained 85.44\% accuracy using the Roshambo dataset and 65.18\% for the EMG subset of sensor fusion dataset.  The classification performance for this method was found to be higher than the ones reported in the literature by using reservoirs with fixed weights in the case of the Roshambo dataset and by using spiking multilayer perceptron (MLP) and spiking convolutional neural network (CNN) for the sensor fusion dataset. This illustrates the high performance of event-based sensing of electrophysiological signals and low power budget and transmission bandwidth requirement.  The 
Spike Encoding \& Evaluation Baseline 
seems to be an excellent reference to which novel neuromorphic SNN classifiers can be compared.

\begin{table}[]
\begin{tabular}{llll}
\hline
Dataset & Ref & Method & Accuracy \\ \hline
\multirow{7}{*}{Roshambo} & \multirow{3}{*}{\cite{Ma2020AICAS}} & \begin{tabular}[c]{@{}l@{}}Reservoir with SVM\\ based readout\end{tabular} & 73.2 \% \\ \cline{3-4} 
 &  & \begin{tabular}[c]{@{}l@{}}Reservoir with\\ spike-rate distance\\ based readout\end{tabular} & 77.0 $\pm$ 0.5\% \\ \cline{3-4} 
 &  & \begin{tabular}[c]{@{}l@{}}Reservoir with STDP \\ based readout\end{tabular} & 60.0 $\pm$ 1.4\% \\ \cline{2-4} 
 & \multirow{2}{*}{\cite{Ma2020JETCAS}} & \begin{tabular}[c]{@{}l@{}}Reservoir with \\spike-rate distance \\ based readout\end{tabular} & 85.28\% \\ \cline{3-4} 
 &  & \begin{tabular}[c]{@{}l@{}}Reservoir with SVM \\ based readout\end{tabular} & 75\% \\ \cline{2-4} 
 & \multirow{2}{*}{\begin{tabular}[c]{@{}l@{}}\textbf{This} \\ \textbf{work}\end{tabular}} & \begin{tabular}[c]{@{}l@{}}Spike Enc. \& Eval. \\Baseline with SVM\\ based readout\end{tabular} & 85.44  $\pm$ 0.75\% \\ \cline{3-4} 
 &  & \begin{tabular}[c]{@{}l@{}}Reservoir (CRITICAL) \\with SVM \\ based readout\end{tabular} & \textbf{88.00 $\pm$ 0.29\%} \\ \hline
\multirow{6}{*}{\begin{tabular}[c]{@{}l@{}}Sensor\\ fusion\\ (EMG)\end{tabular}} & \multirow{4}{*}{\cite{Ceolini2020Frontiers}} & MLP & 67.2 $\pm$ 3.6\% \\ \cline{3-4} 
 &  & CNN & 68.1 $\pm$ 2.8\% \\ \cline{3-4} 
 &  & Spiking MLP & 53.6 $\pm$ 1.4\% \\ \cline{3-4} 
 &  & Spiking CNN & 55.7 $\pm$ 2.7\% \\ \cline{2-4} 
 & \multirow{2}{*}{\begin{tabular}[c]{@{}l@{}}\textbf{This} \\ \textbf{work}\end{tabular}} & \begin{tabular}[c]{@{}l@{}}Spike Enc. \& Eval. \\Baseline with SVM\\ based readout\end{tabular} & 65.07  $\pm$ 3.2\% \\ \cline{3-4} 
 &  & \begin{tabular}[c]{@{}l@{}}Reservoir (CRITICAL) \\with SVM \\ based readout\end{tabular} & \textbf{70.60 $\pm$ 3.65\% }\\ \hline
\end{tabular}
\caption{Comparison of hand gesture classification accuracy with benchmark classification accuracy reported in the literature for the Roshambo dataset and EMG subset of sensor fusion dataset. Note that we use a reservoir with 320 neurons for the 3 class problem, and 2048 neurons for 5 class problem. While ``reservoir" is used as a common term, the methodology for the initalization and the plasticity rules within them widely differ between all the mentioned methods from the literature. }
\label{tab:benchmark_lit}
\end{table}

The architecture shown in Figure~\ref{fig:pipeline_with_reservoir} uses a spiking reservoir with bio-inspired topology from \cite{Balafrej2020PCRITICALAR} for features extraction. With plasticity in the reservoir performance improves further. We implemented the CRITICAL auto-regulation synaptic plasticity in synapses emerging from excitatory neurons.
The classification performance for both datasets is reported in Table \ref{tab:classifier_comp}, and compared in Table \ref{tab:benchmark_lit} with state of the art systems.
The reservoir's performance are compared between fixed weights and with that of a reservoir with CRITICAL plasticity.

Interestingly, the reservoir's performance with fixed weights was lower than that of the 
Spike Encoding \& Evaluation Baseline with SVM readout for both datasets, indicating information loss in the reservoir's activity, possibly due to chaos. 
This highlights the importance of using a reference system like our proposed Spike Encoding \& Evaluation Baseline to challenge neural architectures.
Moreover, the LDA classifier's performance was higher for the fixed weight reservoir than the 
Spike Encoding \& Evaluation Baseline, which validates the proposition that a reservoir projects the signals to a linearly separable hyper-dimensional vector space.

The reservoir performance with regulated activity was found to be more than 88\% (320 reservoir neurons) for the Roshambo dataset and 70.6\% (2048 reservoir neurons) for the sensor fusion dataset. Importantly, the performance figure is higher than fixed weight reservoir \cite{Ma2020AICAS, Ma2020JETCAS} for the Roshambo dataset, and the state-of-the-art CNN \cite{Ceolini2020Frontiers} for sensor fusion dataset as shown in Table \ref{tab:benchmark_lit}. Hence, the initial hypothesis of the reservoir's increased performance at the edge of chaos is corroborated.

\begin{figure}[h]
  \centering
  \includegraphics[width=\linewidth]{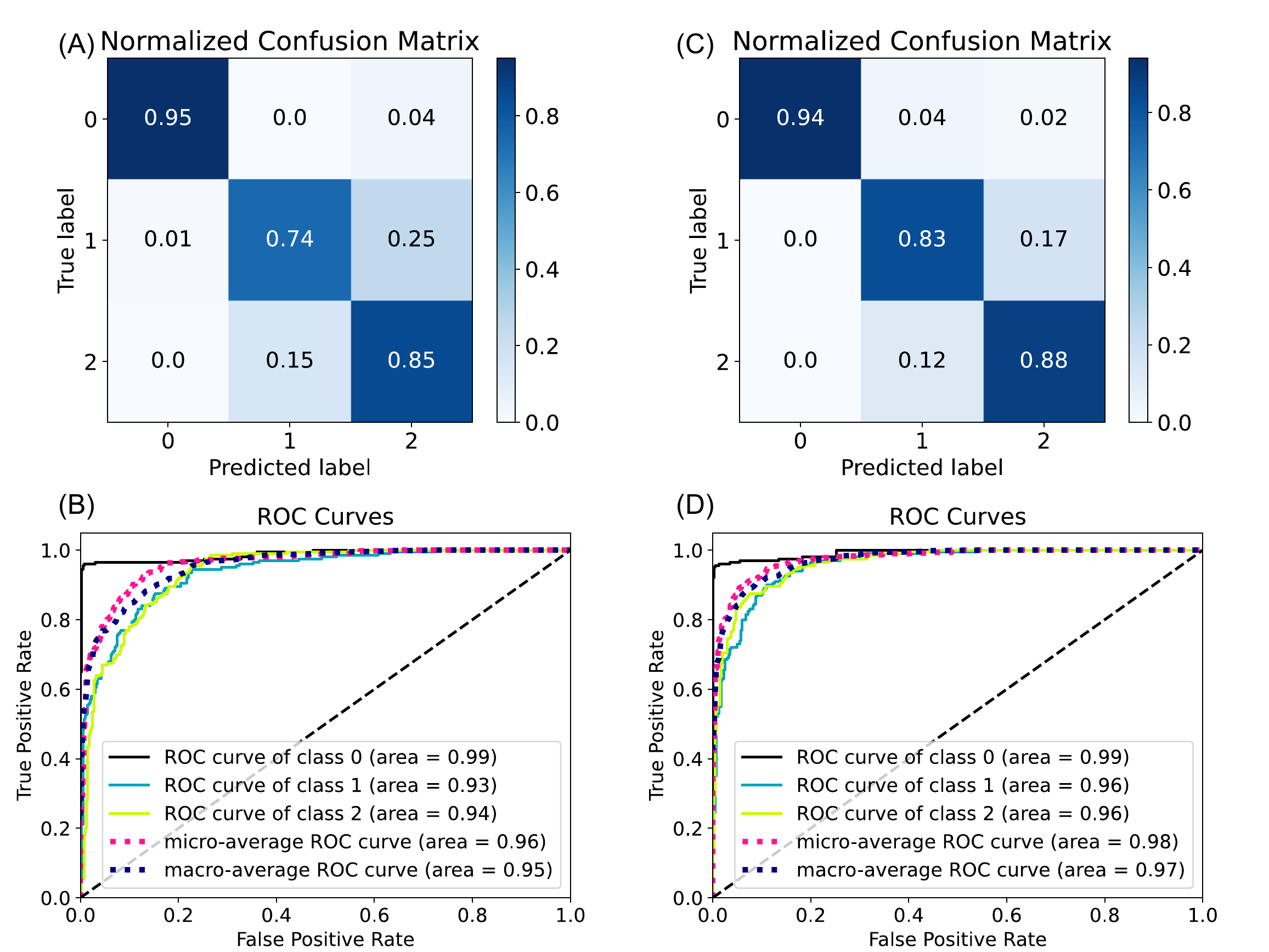}
  \caption{Confusion matrix and ROC curve for Roshambo dataset, where class 0, 1, 2 are 'rock', 'paper', and 'scissor'. (A) and (B) are respectively the confusion matrix and ROC curve for 
the Spike Encoding \& Evaluation Baseline method, (C) and (D) are respectively the confusion matrix and ROC curve of the reservoir of 320 neurons, CRITICAL plasticity, and SVM based readout}
  \Description{Roshambo dataset, where class 0,1,2 are 'rock', 'paper', and 'scissor'. (A) and (B) are the confusion matrix and ROC curve for  
the Spike Encoding \& Evaluation Baseline, (C) and (D) are confusion matrix and ROC curve of the reservoir of 320 neurons, CRITICAL plasticity, and SVM based readout}
  \label{fig:confusion_roshambo}
\end{figure}

\begin{figure}[h]
  \centering
  \includegraphics[width=\linewidth]{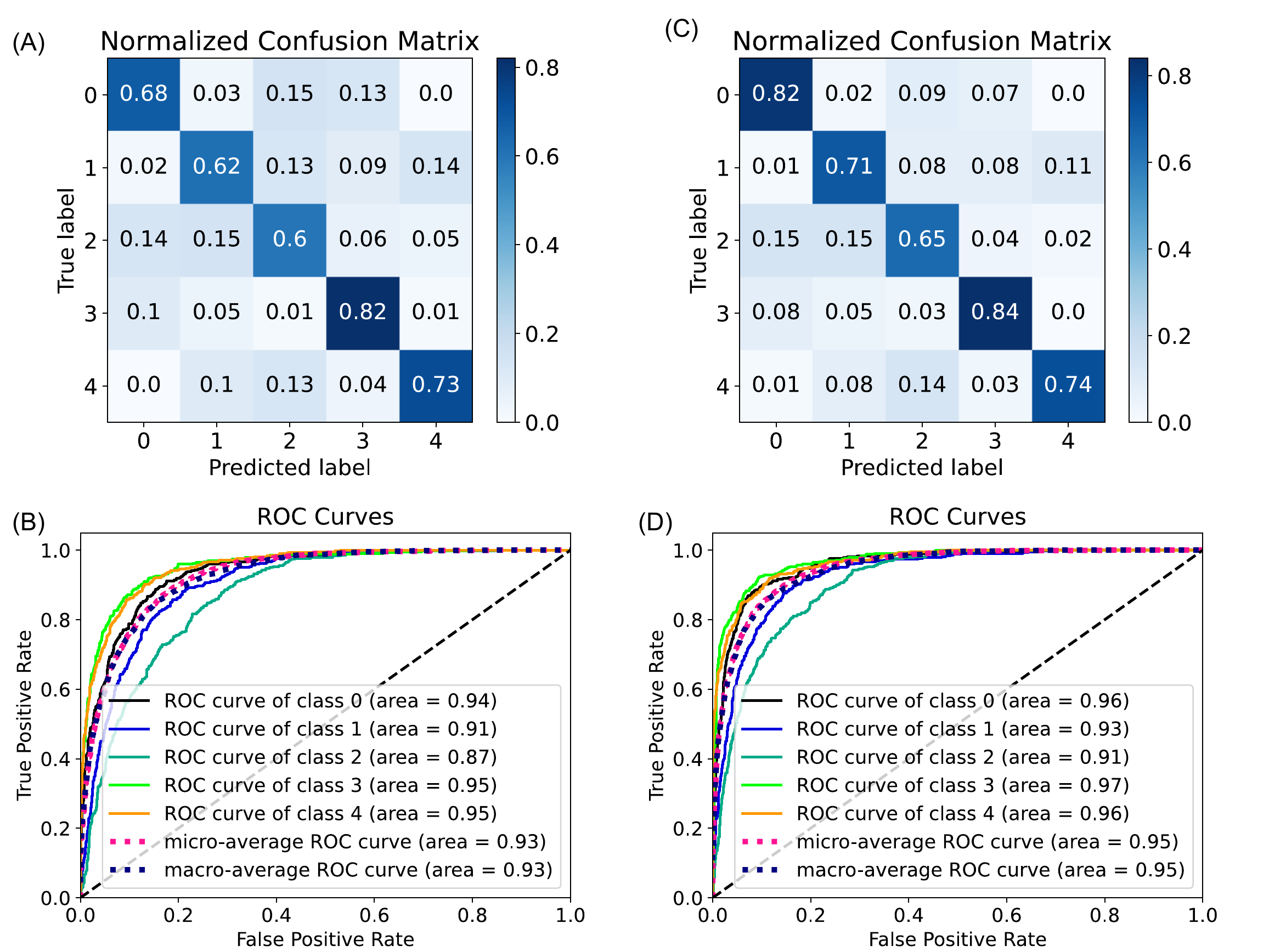}
  \caption{Confusion matrix and ROC curve for sensor fusion dataset, where class 0, 1, 2, 3, 4 are 'pinky', 'elle', 'yo', 'index', and 'thumb'. (A) and (B) are respectively the confusion matrix and ROC curve for  
the Spike Encoding \& Evaluation Baseline method, (C) and (D) are respectively the confusion matrix and ROC curve of reservoir of 2048 neurons, CRITICAL plasticity and SVM based readout}
  \label{fig:confusion_5_class}
\end{figure}

The confusion matrix and receiver operating characteristic (ROC) curve for  
the Spike Encoding \& Evaluation Baseline are shown in Figures \ref{fig:confusion_roshambo}(A) and \ref{fig:confusion_roshambo}(B) for the Roshambo dataset. The ROC curves are used to study the tradeoff between specificity (1 - false-positive rate) and sensitivity (true positive rate), with a classifier closer to the top left indicating better performance and area under curve (AOC) equal to one. AOC score is computed for each class and averaged irrespective of individual class size for computing the macro-average of the AOC. For computing the micro-average, the averaging is weighted by class size. The micro-average and macro-ave
rage of the AOC are almost identical indicating class balance and uniform performance across dataset.
The recognition for the correct class reaches up to 98\% in the case of a regulated reservoir-based approach. Lower accuracy is seen for differentiating movement with similar muscle pathways, i.e., paper and scissor. The confusion matrix and ROC curve for SVM readout performed on encoded data are shown in Figures \ref{fig:confusion_5_class}(A) and \ref{fig:confusion_5_class}(B) for the sensor fusion dataset. The accuracy for the correct class reaches up to 80\%. Designing a system with commands associated with a similar muscle pathway, or low accuracy of differentiability should be avoided.

\begin{figure*} 
\begin{multicols}{2}
{
  \centering
  \includegraphics[width=\linewidth]{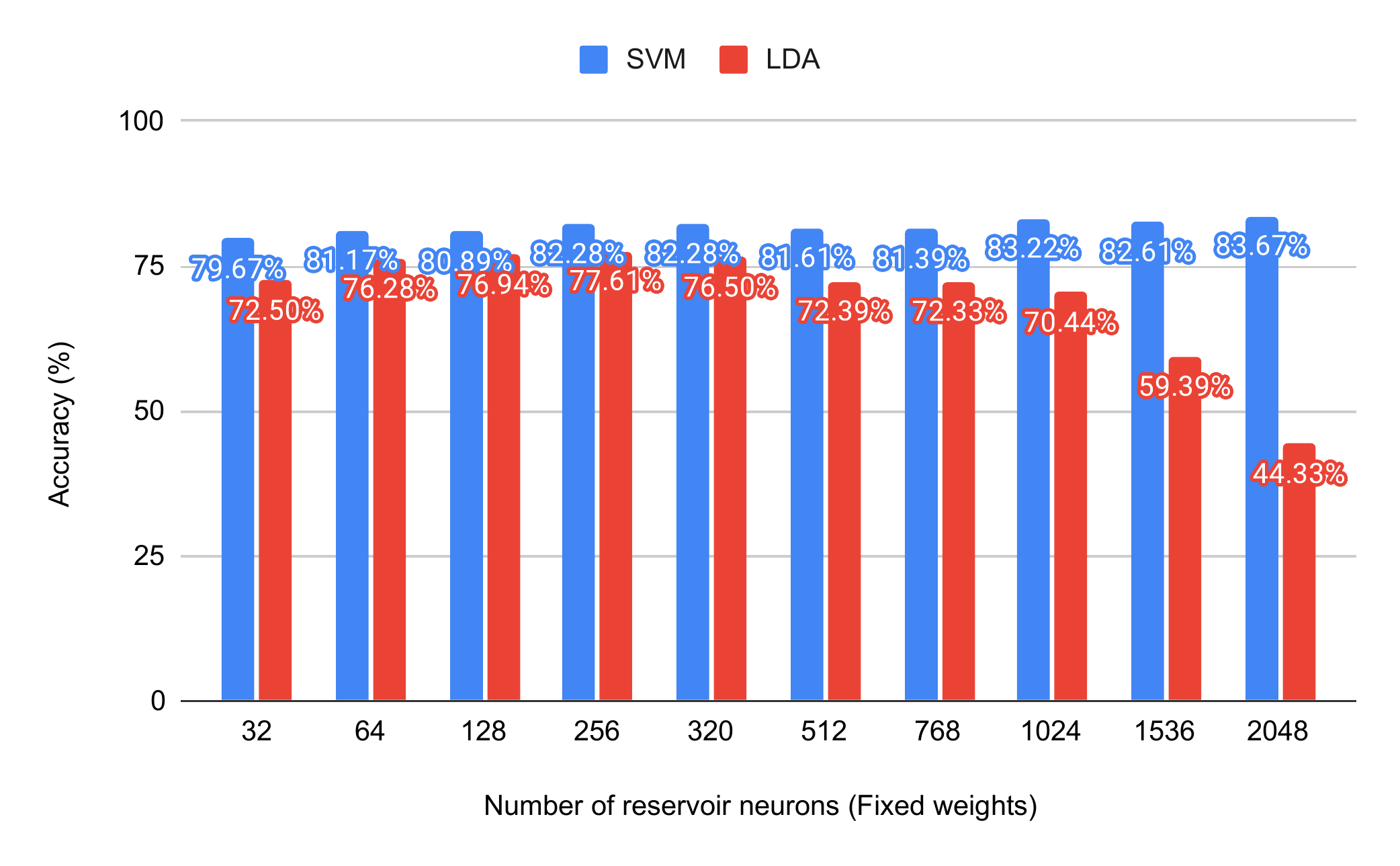}
  \caption{Classification accuracy for the Roshambo dataset plotted against the number of reservoir neurons. The weights of reservoir are fixed (uniform distribution between 0 and 0.25), and SVM or LDA is used in the readout layer.}
  \Description{Classification accuracy for the Roshambo dataset plotted against the number of reservoir neurons. The CRITICAL learning rule is implemented in recurrent connections, and SVM or LDA is used in the readout layer.}
  \label{fig:chart_roshambo_fixed}
}

{
  \centering
  \includegraphics[width=\linewidth]{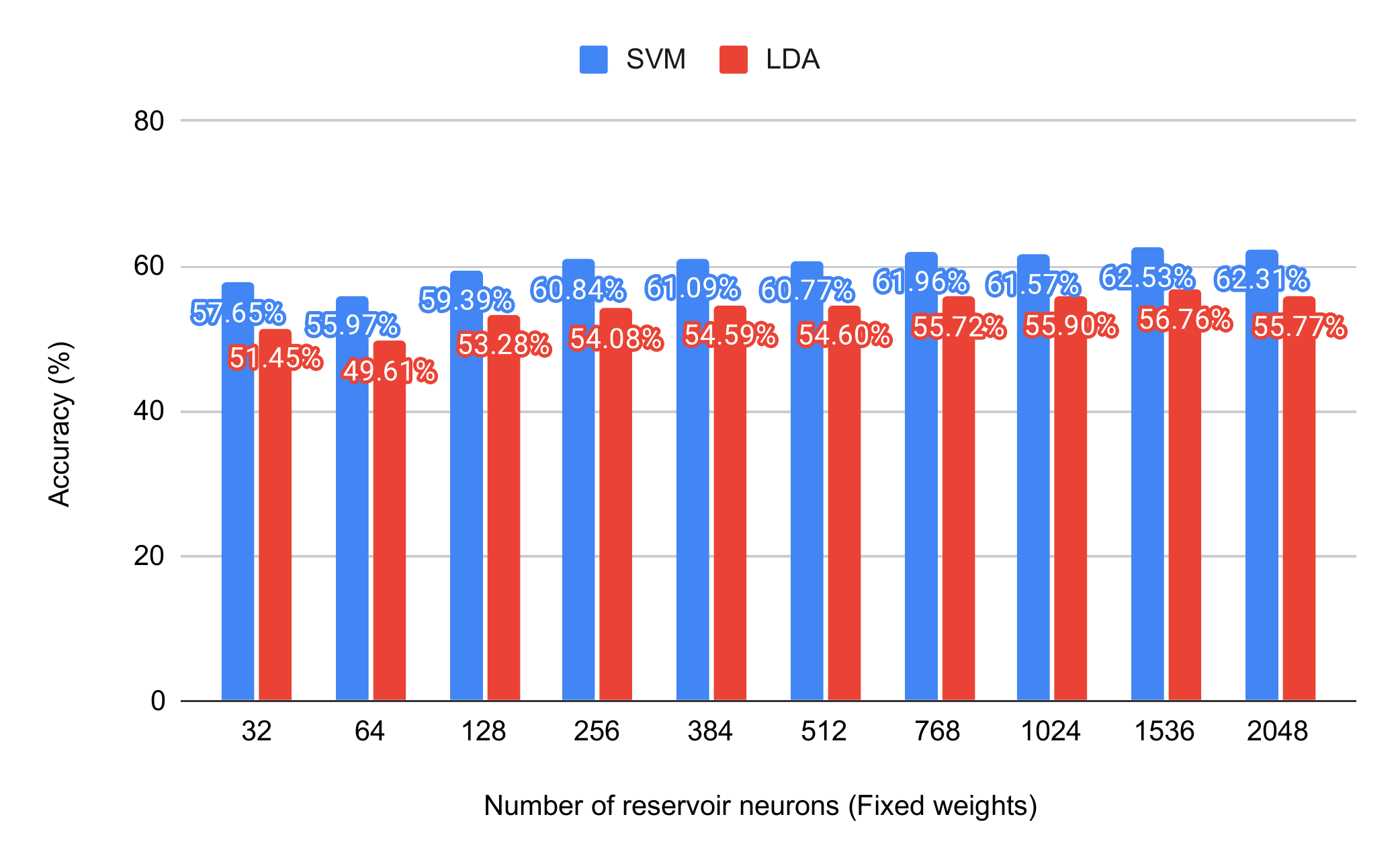}
  \caption{Classification accuracy for the Sensor fusion dataset (5 class) plotted against the number of reservoir neurons. The weights of reservoir are fixed (uniform distribution between 0 and 0.25), and SVM or LDA is used in the readout layer.}
  \Description{Classification accuracy for sensor fusion dataset (5 class) plotted against the number of reservoir neurons. The CRITICAL learning rule is implemented in recurrent connections, and SVM or LDA is used in the readout layer.}
    \label{fig:chart_5_class_fixed}
}
{
  \centering
  \includegraphics[width=\linewidth]{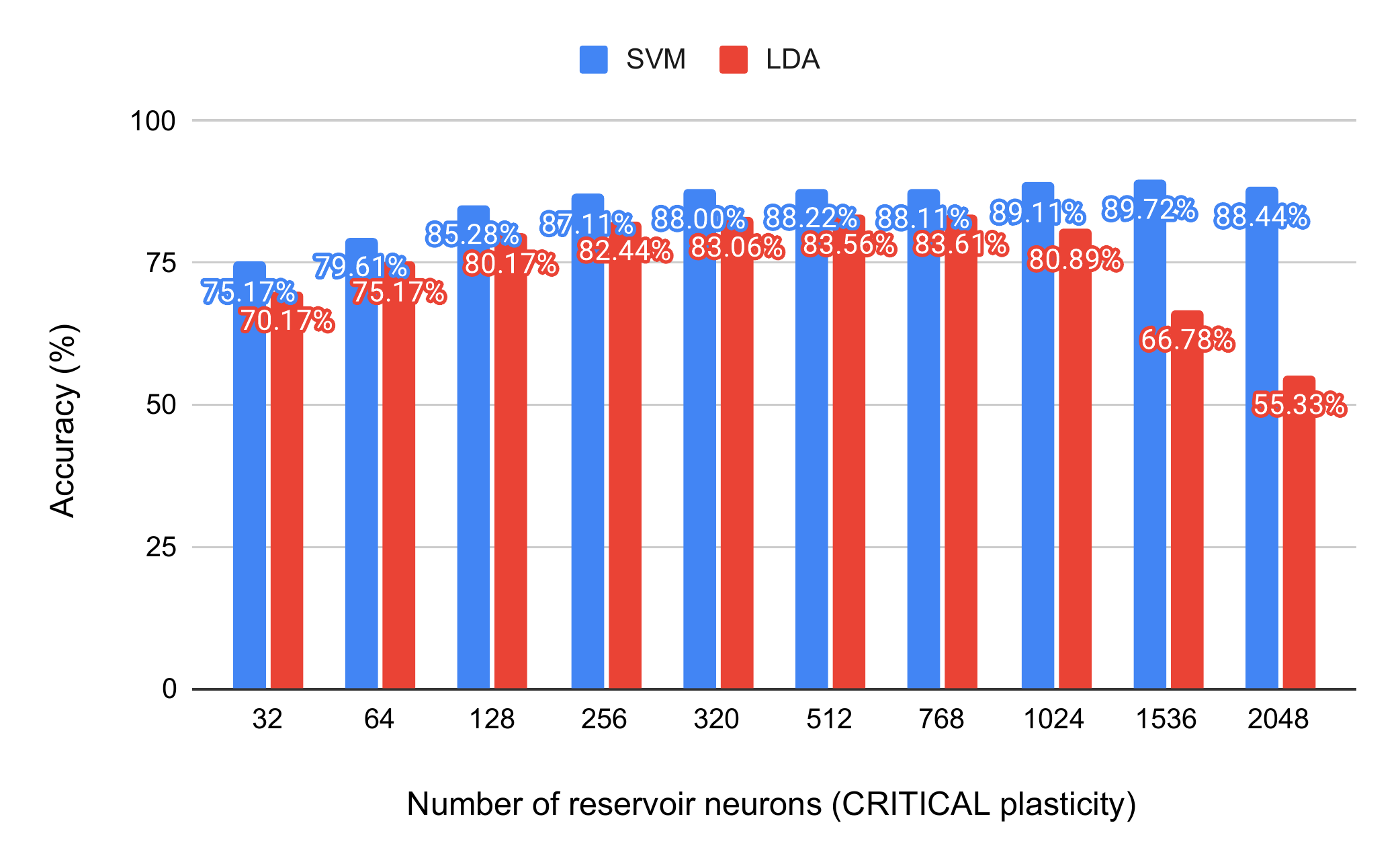}
  \caption{Classification accuracy for the Roshambo dataset plotted against the number of reservoir neurons. The CRITICAL learning rule is implemented in recurrent connections, and SVM or LDA is used in the readout layer.}
  \Description{Classification accuracy for the Roshambo dataset plotted against the number of reservoir neurons. The CRITICAL learning rule is implemented in recurrent connections, and SVM or LDA is used in the readout layer.}
  \label{fig:chart_roshambo}
}
{
  \centering
  \includegraphics[width=\linewidth]{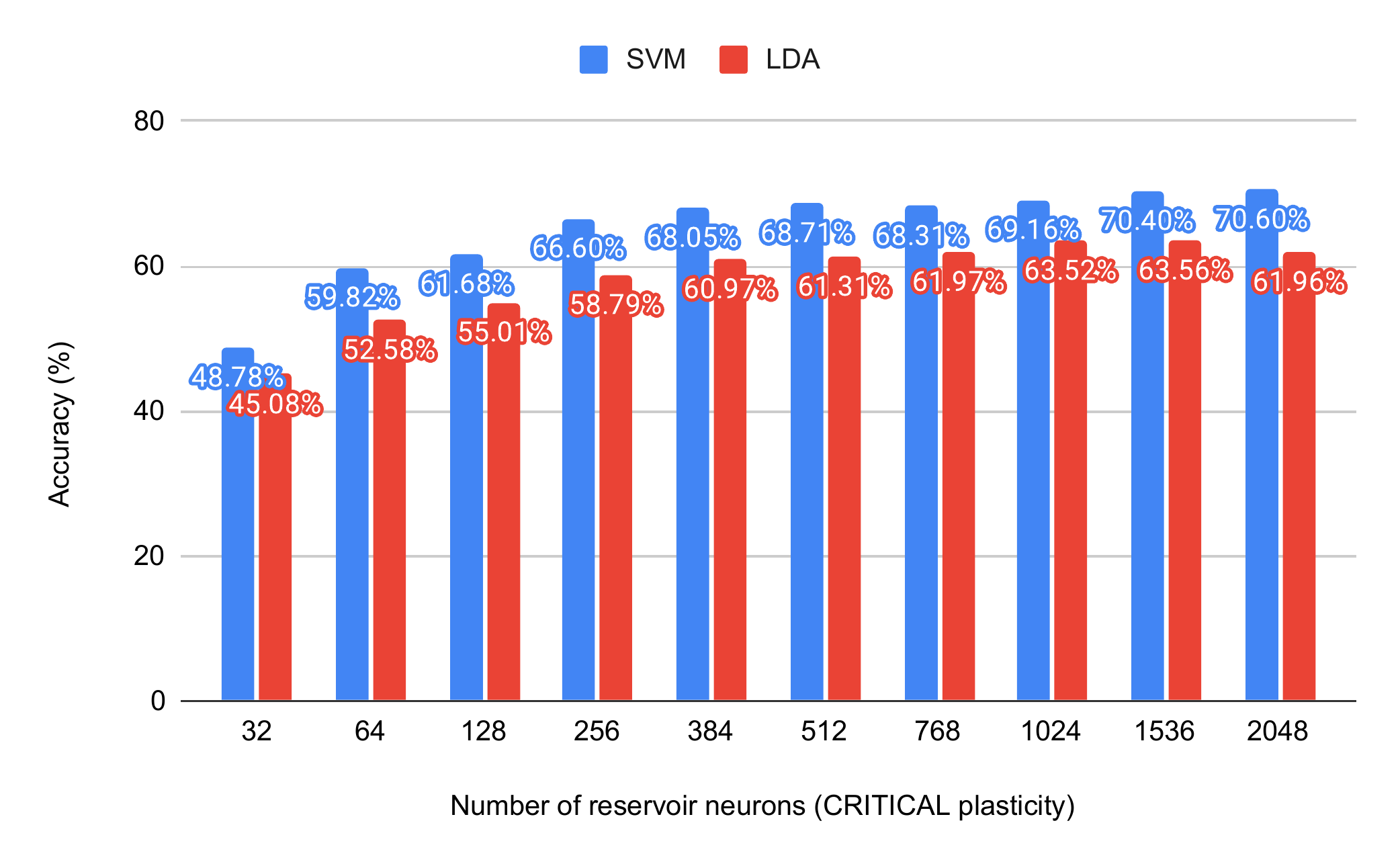}
  \caption{Classification accuracy for sensor fusion dataset (5 class) plotted against the number of reservoir neurons. The CRITICAL learning rule is implemented in recurrent connections, and SVM or LDA is used in the readout layer.}
  \Description{Classification accuracy for sensor fusion dataset (5 class) plotted against the number of reservoir neurons. The CRITICAL learning rule is implemented in recurrent connections, and SVM or LDA is used in the readout layer.}
    \label{fig:chart_5_class}
    }
\end{multicols}
\end{figure*}

The classification performance was plotted against the number of neurons for reservoir with fixed weights and regulated reservoir in Figures \ref{fig:chart_roshambo_fixed} and  \ref{fig:chart_roshambo} and Figures \ref{fig:chart_5_class_fixed} and \ref{fig:chart_5_class} for Roshambo and sensor fusion dataset, respectively. The classification performance increases with the number of neurons consistently for the sensor fusion dataset  
with the SVM-based reservoir's readout. For the Roshambo dataset, the best performance of 89.72 \% was observed for 1536 neurons and activity regulation. This highlights the importance of selecting the size of reservoir for complexity of task (number of classes), and size of dataset. 
A degradation of performance was observed for the LDA classifier while increasing the number of neurons beyond a limit. This highlights the limitation of the simple readout classifier, and need for a more sophisticated classifier like SVMs. 

\section{Conclusion and future scope}

Sensing and processing electrophysiological data is one of the key applications of AI deployment on edge devices. Spike encoding is an important step in the neuromorphic computing pipeline, and it is crucial to ensure that important information for classification is preserved. It is also important to evaluate the efficiency of encoding before passing the information to a SNN to establish a baseline for comparing different SNN topologies and plasticity mechanisms. The high performance of the simple machine learning classifier
(Spike Encoding \& Evaluation Baseline) applied to raw spike converted data in the absence of any signal processing, and the feature extraction method validates the proposition of using event-based sensors to measure biopotential signals. The regulated reservoir of spiking neurons operating with a branching factor of one performs better than the state-of-the-art CNN \cite{Ceolini2020Frontiers}, which further validates reservoir computing's applicability for processing time-series data when operating at the edge of chaos. The reservoir performance with regulated activity reached up to 89.72\% for the Roshambo EMG dataset \cite{donati_elisa_2019_3194792} and 70.6\% for the EMG subset of sensor fusion dataset \cite{ceolini_enea_2020_3663616}. 

In a real-world deployment, essentially, a null class should also be present to avoid the actuator command in the resting state. As the availability of open-source data is limited, such a system could not be trained and evaluated in this study. In the future, we plan to incorporate more gestures along with the resting-state class. Finally, the readout layer, although efficient, is not biologically plausible. In the future, we will explore more biologically plausible readout classifiers in the context of life-long learning.


\section{Author Contributions}

J.R., Y.B., F.A., and D.D. contributed in formulating the study. N.G. designed and performed the experiments and derived the models. N.G., I.B., F.A., and Y.B. analysed the data. N.G. took the lead in writing the manuscript. All authors provided critical feedback and helped shape the research, analysis and manuscript.

\begin{acks}
We acknowledge financial supports from the EU: ERC-2017-COG project IONOS (\# GA 773228) and CHIST-ERA UNICO project. This work was also supported by Natural Sciences and Engineering Research Council of Canada (NSERC) and Fond de Recherche du Québec Nature et Technologies (FRQNT).
\end{acks}




\bibliographystyle{ACM-Reference-Format}
\bibliography{sample-sigconf}


\end{document}